\documentclass[useAMS,usenatbib,epsfig, a4paper]{mn2e}
\usepackage{epsfig}
\usepackage{times}
\usepackage{amsmath, amssymb}

\newcommand{\be}{\begin{equation}}
\newcommand{\ee}{\end{equation}} 
\newcommand{\eei}{\end{equation}\indent\indent}
\newcommand{\bc}{\begin{center}}
\newcommand{\ec}{\end{center}}
\newcommand{\ber}{\begin{eqnarray}}
\newcommand{\eer}{\end{eqnarray}}
\newcommand{\ba}{\begin{array}}
\newcommand{\ea}{\end{array}}

\newcommand{\ls}{{\cal L}}

\newcommand{\lsim}{\,\raise 0.4ex\hbox{$<$}\kern -0.8em\lower 0.62ex\hbox{$\sim$}\,}
\newcommand{\gsim}{\,\raise 0.4ex\hbox{$>$}\kern -0.7em\lower 0.62ex\hbox{$\sim$}\,}

\def\case#1/#2{\textstyle\frac{#1}{#2} }
\newcommand{\bra}[1]{\left(#1\right)}
\newcommand{\bras}[1]{\left[#1\right]}

\newcommand{\reff}[1]{(\ref{#1})}
\newcommand{\Mpc}{\text{Mpc}}
\newcommand{\zmax}{z_{\textrm{max}}}
\newcommand{\wv}{{\bf w}}
\newcommand{\model}{{\bf m}}
\newcommand{\data}{{\bf D}}

\title[Reconstructing dark energy with maximum entropy]{Reconstructing the history of dark energy using maximum entropy}
\author[Zunckel and Trotta]{Caroline~Zunckel\thanks{E-mail address: {\tt clz@astro.ox.ac.uk}} and Roberto~Trotta\thanks{E-mail address: {\tt rxt@astro.ox.ac.uk}}\\
Oxford University, Astrophysics,  Denys Wilkinson Building, Keble
Road, OX1 3RH, United Kingdom}
\date{\today}
\begin{document}
\maketitle
\begin{abstract}
We present a Bayesian technique based on a maximum entropy method
to reconstruct the dark energy equation of state $w(z)$ in a
non--parametric way. This MaxEnt technique allows to incorporate
relevant prior information while adjusting the degree of smoothing
of the reconstruction in response to the structure present in the
data.

After demonstrating the method on synthetic data, we apply it to
current cosmological data, separately analysing  type Ia
supernovae measurement from the HST/GOODS program and the first
year Supernovae Legacy Survey (SNLS), complemented by cosmic
microwave background and baryonic acoustic oscillations data. We
find that the SNLS data are compatible with $w(z) = -1$ at all
redshifts $0 \leq z \lsim 1100$, with errorbars of order $20\%$
for the most constraining choice of priors. The HST/GOODS data
exhibit a slight (about $1\sigma$ significance) preference for
$w>-1$ at $z\sim 0.5$ and a drift towards $w>-1$ at larger
redshifts, which however is not robust with respect to changes in
our prior specifications. We employ both a constant equation of
state prior model and a slowly varying $w(z)$ and find that our
conclusions are only mildly dependent on this choice at high
redshifts.

Our method highlights the danger of employing parametric fits for
the unknown equation of state, that can potentially miss or
underestimate real structure in the data.

\end{abstract}
\begin{keywords}
Cosmology: dark energy -- methods: data analysis, Bayesian
techniques
\end{keywords}
\section{Introduction}
With the confirmation of the accelerated expansion of the universe
\citep{Perlmutter:1998np,Lange:2000iq,Hoekstra:2002xs,Riess:2004nr,Cole:2005sx,Astier:2005qq,
Spergel:2006hy,Riess:2006fw} comes the inference that the cosmic
dynamics are today dominated by a component that competes
\emph{against} gravitational collapse of matter and thus must have
negative pressure \citep{Frieman:1995pm,Peebles:2002gy}: this has
been dubbed dark energy. All observations are presently compatible
with dark energy being in the form of Einstein's cosmological
constant $\Lambda$, a new form of matter--energy with equation of
state (EOS) $w=\rho/p=-1$. However, it has been shown that an
equation of state which changes with redshift, $w(z)$, can mimic a
cosmological constant and fit the current data \emph{if the
parameterization of $w$ is assumed to be a constant}
\citep{Linder:2004rm,Simpson:2006bd}. At the same time, an
explanation based on the cosmological constant still suffers from
the so--called ``coincidence'' and ``fine tuning'' problems, and
it remains unclear whether selection effects of the kind embodied
by anthropic arguments can offer a solution
\citep{Starkman:2006at}.

Determining whether dark energy is constant in time or has
dynamical properties is one of the most pressing outstanding
questions in cosmology, as witnessed by the multiplication of
observational proposals trying to elucidate the question (see e.g.
\cite{Trotta:2006gx} for an overview). We are thus led to question
the simple solution of a constant $w(z)$, especially in view of
the fact that several recent works have highlighted the dangers of
fitting current data by assuming a specific parameterization for
the dark energy EOS \citep{bassett-2004-617, Linder:2004rm}.

The purpose of this paper is to explore the use of Bayesian
statistical techniques based on a maximum entropy method to
investigate the time dependence of dark energy by imposing minimal
assumptions on the functional form of $w(z)$. Whenever external
(prior) information is used, its impact is clearly expressed by
our formalism, making the reconstruction totally transparent. To
do so we require information about the expansion of the universe.
The quality and quantity of observational data of cosmological
relevance is rapidly increasing: type 1a supernova (SNIa) (see
e.g.~\cite{Riess:2004nr,Astier:2005qq}) can be calibrated to serve
as standard candles (see~\cite{Nugent:2006pw} for a recent
proposal of using type II-P supernovae instead). Supernovae type
Ia observations can thus measure the luminosity distance as a
function of redshift, $D_L(z)$,
 \be  \label{eq:1} D_L(z)= \frac{c}{100 h} (1+z)
 \int^{z}_{0} \frac{1}{H(x)} dx \quad[\Mpc],
 \ee
where the present--day Hubble constant is $H_0 = 100 h$ km/s/Mpc,
$c$ is the speed of light in km/s and the redshift dependent
Hubble function $H(z)$ can be expressed in terms of the
present--day matter--energy content of the Universe as (here and
in the following we assume a spatially flat Universe)
 \ber H^2(z)&=&
 \bra{1-\Omega_m-\Omega_{DE}}(1+z)^2+
 \Omega_m (1+z)^3 \nonumber \\
 &+&\Omega_{DE} \exp \bras{3\int^{z}_0 \frac{1+w(z)}{1+z} d z} . \label{Hz} \eer
Here, $\Omega_m$ is the the density of matter in units of the
critical density today and $\Omega_{DE}$ is the (present--day)
dark energy density in units of the critical density. The dark
energy density time evolution is determined by its equation of
state (EOS) $w(z)$,
 \be \rho_{DE}(z)=\rho_{DE}(0) \exp
 \bras{3\int^{z}_0 \frac{1+w(z)}{1+z} d z}. \label{rho}
 \ee
Observations of the luminosity--distance find a powerful geometric
complement in the use of  ``standard rulers'' such as the position
of the acoustic peaks in the cosmic microwave background (CMB)
power spectrum \citep{Bennett:2003bz,Spergel:2006hy} and the
(transversal) baryon acoustic oscillation (BAO) signature recently
measured in the galaxy matter power spectrum
(\cite{Eisenstein:2005su,Cole:2005sx}). Such data can be used to
constrain the angular diameter distance $D_A(z)$
 \be D_A(z)= (1+z)^2 D_L(z).
 \label{D_A} \ee

Apart from its geometrical impact on the angular diameter and
luminosity distance relations, the properties of dark energy also
influence the growth of structures and can therefore be
constrained through weak lensing (see
e.g.~\cite{Hu:2002rm,Hoekstra:2005cs,Jarvis:2005ck}) and cluster
counts data. In order to constrain $w(z)$ from measurements of
either $D_L(z)$ or $D_A(z)$, we need to perform two derivatives,
as it is evident from \reff{Hz} and \reff{rho}. This is
problematic if we consider the increase in the noise that
accompanies each derivative. In addition to information loss through this
indirect determination of $w(z)$, the current data tends to be
sparse with a large sample variance.

For these reasons, it seems timely and relevant to shift attention
to establishing more powerful statistical methods to extract in
the most efficient and faithful way the information on $w(z)$
contained in present and upcoming large data sets. The development
of a technique that can cope with the patchy distribution in
redshift space while making minimal assumptions on the time
properties of dark energy is the logical next step towards
improving our understanding of dark energy. In this paper we apply
a modified Maximum Entropy (MaxEnt) technique that has been
diversely used to successfully reconstruct images and spectra
under unfavourable conditions (for applications to astrophysical
problems, see
e.g.~\cite{bridle-1998,Marshall:2001ax,Maisinger:2003hd}). With a
firm basis in probability theory, the method can be tailored to
the needs of dark energy reconstruction from present data. Our
application of MaxEnt aims at reconstructing the dark energy EOS
while minimizing our assumptions regarding the form of $w(z)$.

This paper is organized as follows. We firstly outline the
statistical framework of our technique in section \ref{sec:stat}.
We then proceed to test our reconstruction method on synthetic
data in section \ref{sec:demo} and then apply it to present
cosmological data in section \ref{sec:application}. We offer our
conclusions in section \ref{sec:conclusions}.

\section{Maximum entropy reconstruction technique}
\label{sec:stat}
\subsection{Motivation}
When attempting to constrain the nature of dark energy, a
procedure common in the literature is to Taylor expand the
quantity $\rho_{DE}(z)$ or $w(z)$ around $z=0$, then constraining
the expansion coefficients through the data
\citep{efstathiou-1999,huterer-2001,Weller:2001gf}. An example of
such a parameterization that is commonly employed is
 $w(z) =  w_0 + (1- a) w_1$ where  $a=\frac{1}{(1+z)}$ \citep{Chevallier:2000qy, linder-2003-90}.
Alternatively, one might prefer to parameterize the
time--dependence of the EOS using some smooth function (such as
the ones used in \cite{Dick:2006ev}), that is hoped will
encapsulate the essential features of the dynamics one wishes to
constrain.  Both procedures are not free from the risk of giving
misleading results, since they impose artificial assumptions on
the form of the EOS, which often have no basis in any physical
mechanism. \cite{bassett-2004-617} highlight the dangers of
implementing such parameterizations. Moreover, this will only be
sensitive to departures from a constant density within a
restricted set of models \citep{Dick:2006ev}.

\cite{Huterer:2002hy} introduced a principle component analysis
(PCA) of the function $w(z)$, with the aim of adopting a
parameterization appropriate to the data sets used (see also
\cite{Dick:2006ev,Simpson:2006bd} for related issues). These PCA
modes are argued to form a natural basis in which to characterize
dark energy evolution and by using only the first few
well--determined eigenvectors in the reconstruction one tries to
exclude noisy modes and thereby gain accuracy in the
reconstruction. However, the method has the disadvantage of
introducing an ill--controlled bias at high redshifts, i.e.~the
removal of strongly oscillating (and noisy) modes may mislead one
to the conclusion that the EOS reverts to the fiducial model at
large redshift with artificially small error bars. While we
recognize the merits of the PCA method, we wish to improve on it
in this last respect by making fully explicit the assumptions that
will control the behaviour of the reconstructed $w(z)$ at large
redshifts.

The MaxEnt technique we employ has parallels with the well known
maximum likelihood (ML) approach, but introduces new features
ensuring that in the case where insufficient information is
available the most likely distribution is the most uniform, i.e.
the one with \emph{maximum entropy} (or minimum information
content. For an overview of the connection between entropy and
information content, see e.g. \cite{Trotta:2005ar,Kunz:2006mc}).
In other words, where ML merely maximizes the likelihood, often
unnecessarily overfitting the noise in the data, MaxEnt seeks the
optimum trade--off between a smooth, maximally entropic
distribution and the rough distribution mapped out by the data.
The most important characteristic is that the MaxEnt method is
auto--regulating, i.e. the amount of smoothness (or raggedness) in
the reconstruction is consistently determined through the data
themselves (see section \ref{sec:alpha} below). In our Bayesian
perspective, extra information coming from prior beliefs or
theoretical prejudice can be naturally incorporated in the
reconstruction via the MaxEnt prior. As we show below, this gives
MaxEnt the power to cope with situations where the dimensionality
of the parameter space potentially exceeds the number of data
points, a difficult reconstruction problem that is ill--defined
under ML techniques. This feature clearly makes MaxEnt highly
applicable to the case of dark energy reconstruction.
\subsection{The MaxEnt formalism}

The task at hand is to determine the EOS of dark energy from
sparse data on $D_L$, $D_A$ and, to a limited extent, on $H(z)$
(as encapsulated by today's detection of BAO). We consider a
piece--wise constant $w(z)$ in $N$ bins out to a maximum redshift
$\zmax$. Let $w_j$ be the value of the EOS in the $j$--th bin, $1
\leq j \leq N$. In analogy with the treatment given in
\cite{Skilling}  for the case of image reconstruction, we gather
all the EOS bins values in an ``image vector'' $\wv$. We seek to
determine the posterior distribution of $\wv$ given the observed
data $D$, $Pr(\wv \mid \data)$. This is obtained through Bayes'
theorem as
  \be Pr(\wv \mid \data)=\frac{Pr(\data \mid
 \wv) Pr(\wv)}{Pr(\data)} .\label{posterior}
  \ee
  The quantity $Pr(\wv)$ is the prior probability representing all the
information about the distribution $\wv$ before the data $\data$
has been collected; $Pr(\data \mid \wv)$ is the likelihood and
describes the underlying statistical process and $Pr(\data)$ is
the model likelihood (``evidence''), which is relevant for model
selection questions but that is unimportant in this case. We shall
therefore neglect this proportionality constant in the following.
\subsubsection{The MaxEnt prior}
The principle of MaxEnt is employed to determine a prior $Pr(\wv)$
that encapsulates all the external information about $w(z)$ we
wish to specify in the absence of the data. Following
\cite{Skilling}, we adopt the principle that the least biased
model that encodes any given prior information is the one which
maximizes the entropy of the distribution while remaining
consistent with the information. This prior is appealing for its
characteristic of maximising the uncertainty (entropy) of the
distribution thus making minimal assumptions.
The MaxEnt prior
\citep{Skilling} takes the form
 \be Pr\bra{\wv \mid \alpha, \model} =
 \frac{\exp \bra{\alpha S(\wv, \model)}}{Z_S} \label{S} \ee where
$S\bra{\wv,\model}$ is the entropy of $\wv$ relative to the model
$\model$ and $\alpha$ is a regularizing constant. The model
$\model$ defines the image vector to which $\wv$ reverts in the
absence of any data, and as such it defines a measure on the DE
parameter space. The information entropy is analogous to the
thermodynamic entropy in statistical mechanics, which is given by
the logarithm of the number of states by which one can arrive at a
given macroscopic constraint. In the same way, the information
entropy can be described as the logarithm of the numbers of ways
in which one can arrive at a particular $\wv$ in a Poisson process
when starting with the model $\model$ \citep{PhysRevB.41.2380}.
The entropy $S$ for an $N$--dimensional discrete parameter space
is \citep{Skilling}
 \be
 S(\wv, \model) = \sum^{N}_{j=1} \bras{ w_j - m_j - w_j \log
\bra{\frac{w_j}{m_j}}}. \ee The log term is reminiscent of the
Kullback--Leibler divergence between $\wv$ and $\model$, encoding
the amount of information present in $\wv$ with respect to the
model $\model$. In our case, we do not apply the entropy to the
values of $\wv$ directly, but rather to the space of coefficients
of an expansion of $\wv$ in a series of basis functions (that we
choose to be top--hat functions in redshift space, see
section~\ref{sec:paramet} for details). We can think of the
coefficients of the expansion as a series of weights that encode
how much each basis function contributes to the total $w(z)$. We
can then apply the MaxEnt prior on the space of these weights, by
thinking of them as expressing the relative contribution of each
basis function -- in other words, in a phenomenological way we
take the weights to represent relative probabilities for the
presence of each basis function in the final $w(z)$. Below, we
will use the notation $\wv$ as a shortcut to indicate the vector
of weights of the dark energy expansion. The same applies to the
model $\model$, that in the entropy term is represented by its
expansion coefficients in the chosen basis functions.

Evidently, $S(\wv)$ (for a fixed $\model$) is a convex function
which reaches a maximum for $\wv=\model$ with a value $S=0$. Thus
in the absence of any information from the data, the entropy term
reverts to the model. The normalizing partition function for the
entropy is given by
 \be
 Z_S =\int \exp{{\alpha S(\wv)}} \det[g]^{1/2} d^{N} \wv.
\ee
 The measure is defined as the invariant volume $\det g^{1/2}$
of the metric $g$ defined on the space where $g_{ii}=1/w_i$ and
$g_{ij} = 0$ for $i\neq j$ (also known as the Fisher information
matrix). By expanding to second order around the model
$\wv=\model$ (at the maximum $S = 0$), we obtain the partition
function in the Laplace approximation:
 \be Z_S=\bra{\frac{\alpha}{2\pi}}^{N/2}
\ee where $N$ is the number of parameters, in our case the number
of expansion coefficients for $w(z)$.

\subsubsection{The likelihood}
 The likelihood is defined as
of the probability of the \emph{data given the parameters}:
 \be Pr(\data \mid \wv) = \frac{\exp(-
 \ls(\wv))}{Z_{L}}, \label{like} \ee and is the probability that
the observed data $\data$ could have been generated from a given
$\wv$. For data $\data$ subject to Gaussian noise the likelihood
function is
 \be \ls
 (\wv)=\frac{1}{2}(\data - f(\wv))^T[C^{-1}](\data - f(\wv)),
 \label{L} \ee
where $C$ is the data covariance matrix and $f(\wv)$ denotes the
functional dependence of the observable on the DE parameters in
our case, $f=H$ or $f=D_A$ ($D_A$ and $D_L$ being simply related
through the redshift, see~\eqref{D_A}). In the case of independent
data points with uncorrelated noise, the covariance matrix is
diagonal with the non--zero elements being the variances of each
measurement, denoted by $\sigma_i^2$, $i=1,\dots,N_D$. The
normalizing partition function for $\ls$ is
 \be Z_L=\int \exp \bra{\ls (\wv)} d^{N_D} D. \ee
 Using the identity for the normalized probability distribution we
obtain:
 \be Z_L=\frac{(2\pi)^{N_D/2}}{\sqrt{\det[C^{-1}]}}. \ee

\subsubsection{The posterior probability}

From the likelihood in Eq.~\eqref{like} and the prior in
\eqref{S}, we obtain from Bayes' theorem the posterior probability
for the DE parameters $\wv$:
 \be Pr(\wv \mid \data,  \alpha, \model)\propto
 \exp(\alpha S(\wv) - \ls(\wv)).\ee
  Given that $ \ls(\wv)$ is
quadratic in $\wv$ and $S(\wv)$ is a convex function, the above is
well--constrained with the peak of the posterior for $\wv$ being
determined by a competition between $S$ and $\ls$, mediated by the
value of $\alpha$. We thus see that the MaxEnt prior will be
useful in the case where the parameter space dimensionality
exceeds the size of the dataset in that the entropy is
incorporated as a regularization to avoid over--fitting, while
retaining maximum flexibility in the underlying parameterization
of $w(z)$. $S$ penalizes the excess ``structure'' in the data,
with the regularizing parameter $\alpha$ dictating the degree of
this smoothing. The choice of $\alpha$ is thus very important: a
small value of $\alpha$ produces little smoothing and the
likelihood will dominate, resulting in a reconstructed
distribution where the noise might be mistaken as real structure.
Alternatively, too large a value for $\alpha$ leads to information
loss, with the entropic prior overriding the information coming
from the data.
\subsubsection{The regularization parameter $\alpha$}
\label{sec:alpha} In order to select the correct value of $\alpha$
we add it to the hypothesis space as an additional parameter and
let the data select the optimum value. Using once more Bayes'
theorem we have
  \be
 Pr\bra{\alpha \mid \data, \model} \propto Pr\bra{\alpha}
Pr(\data\mid\alpha, \model), \ee
 where $Pr(\alpha)$ is the prior on $\alpha$. The joint posterior
probability is then (up to irrelevant constants)
 \ber \label{eq:jointpost}
 Pr\bra{\wv,\alpha,\data | \model}& \propto&Pr(\alpha)Pr(\wv \mid \alpha) Pr(\data \mid \wv) \nonumber \\
 &=&Pr(\alpha) \frac{\exp \bra{\alpha S(\wv)-\ls(\wv)}}{Z_S(\alpha)
 Z_L}. \eer
We adopt a Jeffreys' prior on $\alpha$, which is flat in $\gamma =
\log\alpha$, reflecting ignorance of the scale of the variable,
within the range $-10 \leq \gamma \leq 10$. This corresponds to
choosing $Pr(\alpha) \propto 1 /\alpha$. In the final inference on
$\wv$ we marginalize over the nuisance parameter $\alpha$, even
though the distribution of $\alpha$ is usually fairly strongly
peaked and thus marginalization is almost equivalent to
maximisation (i.e., simply fixing $\alpha$ to the value of the
peak of the posterior).

\subsubsection{Model specification}
\label{sec:mod}

The joint posterior in Eq.~\eqref{eq:jointpost} is conditional on
the specific choice of model $\model$, to which the reconstructed
$\wv$ defaults in the absence of data. The entropic prior
distribution is introduced to penalize the posterior for
unwarranted complexity. Given that the model is the vector to
which $\wv$ should revert in the absence of data, it must
represent maximal smoothness.  We need to establish what
distribution $\model$ will encompass this in the context of the
EOS of dark energy. In image reconstruction the default model is
usually taken to be a flat surface equal to the mean of the data.
When the data is then included via the likelihood, variation about
this mean is introduced.  In our case, this means choosing a
contant model, $\model=\text{const}$ that is \emph{uniform} in
redshift space. There is however no obvious choice for the
magnitude of this constant. There are various possible choice of
$\model$, reflecting different prior beliefs about the dark energy
EOS. One can thus usefully think of $\model$ as encapsulating a
fiducial, reference model we want to test the observations
against. One possibility is to set $\model = -1$ at all redshifts,
thus representing a cosmological constant. This is recommended if
we are testing for deviations from $w(z) = -1$: if significant
deviations from the model are found in the reconstructed EOS, this
is an indication that the data are informative enough to override
the entropic pull towards the model, and thus that such deviations
are likely to reflect real structure in the data. A more skeptical
attitude towards dark energy might be encapsulated by choosing a
constant model $\model = 0$, which correspond to a pressureless,
dust--like fluid. In this case, if the reconstructed $w(z)$
assumes values below 0, this can be interpreted as a strong
indication for the presence of a fluid with negative pressure,
with data being strong enough to dominate the entropic tendency
for a pure matter Universe. In principle, a theoretical prejudice
in the form of a particular redshift--dependence of $\model$ could
also be implemented easily in the same fashion.

Finally, one can also employ Bayes theorem to take $\model$ into
the joint posterior, by writing
 \be
 Pr(\wv, \data, \alpha, \model) \propto  Pr(\model) Pr(\wv, \data, \alpha | \model)
 \ee
and marginalizing over $\model$ in the left--hand--side, after
specifying a prior over the model space, $Pr(\model)$. In this
work we take all the model vectors to be constant over the whole
redshift range, thus the specification of the model amounts to the
choice of the value of the constant. We restrict our
considerations to the range $-1 \leq \model \leq 0$, and when
performing a marginalization over the model we will take a flat
prior in this range, i.e. $Pr(\model) = \text{const}$.

\subsection{Dark energy parameterization and reconstruction}
\label{sec:paramet}

As motivated above, we decompose $w(z)$ into a weighted sum of
orthogonal functions in redshift space, with the parameters being
given by the weights encoding the amount that each function
contributes to the overall $w(z)$. There are of course several
different meaningful expansion functions such (for example,
principal components, Chebychev functions, etc) but we make use of
the simplest option, decomposing $w(z)$ is into a series of $N$
step--functions $\Phi_i(z)$:
 \be \label{eq:coeff}
 w(z)=-2+\sum^{N}_{i=1} C_i\Phi_i(z) \ee
 where
$\Phi_i(z)=1$ for $z_i-\Delta z/2<z<z_i+\Delta z/2$ and
$\Phi_i(z)=0$ everywhere else.  Since the least stringent limits we will impose on $w(z)$ are $-2\leq w_i \leq 0$, the above ensures that  the expansion
coefficient $C_i$ ($i=1,\dots,N$) are positive numbers, a
necessary requirement for our entropic prior. The parameter space
$\wv$ is thus constructed from the co-efficients $C_i$ themselves, which
are allowed to vary within the range $0 \leq C_i \leq 2$.
An advantage of this piece-wise parameterization of $w(z)$ is that
it will be possible to capture a sharp change in the EOS, provided
the binning is sufficient. In order to capture different features
of the time evolution of the EOS, other expansion functions may be
more appropriate. We experimented with Chebychev functions and
found that their oscillatory behaviour was not helpful in
reconstructing sharp changes in the EOS. Such smooth functions
might be more useful if one wants to test quintessence models
exhibiting a gentle evolution of $w(z)$.

In the bulk of recent analyses the limit $-1\leq \wv \leq 0$ is
imposed; the lower limit stemming from the null energy condition
which must be satisfied for dark energy to be stable
\citep{Alcaniz:2003qy}. Although models of dark energy that allow
$w< -1$ violate the weak energy condition in the context of
general relativity, these ``phantom'' components have been studied
by many authors \citep{Caldwell:1999ew}. There have been claims
that such phantom behaviour is unstable when regarded as a quantum
field theory \citep{Carroll:2003st}. From a phenomenological
perspective, it makes sense to both restrict the range of our
reconstruction to lie within $-1 \leq \wv \leq 0$, but also to
extend the parameter space to values below $\wv = -1$ to check the
stability of the reconstruction. We will thus presents results
also for the case where the equation of state can attain values as
low as $\wv = -2$.
\begin{table}[t]
\begin{center}
\begin{tabular}{ccc}    \hline
\textbf{Parameter}  &   \textbf{Prior} &
\\ \hline
$\Omega_{DE}$  & $0.0\dots1.0$  &  Top--hat
\\
$\Omega_{\kappa}$  & 0  &  Flatness imposed \\
$C_i$ & $0\dots 2$ &   Top--hat
\\
$\gamma=\log(\alpha)$ & $-10 \dots 10$ &  Top--hat
\\
$\model$ & $-1 \dots 0$ & Top--hat \\\hline
\end{tabular}
\end{center}
\caption{Priors on cosmological and nuisance parameters used in
the analysis. We employ a Jeffreys' prior on $\alpha$, i.e. we
take the prior to be flat in $\gamma = \log\alpha$ to reflect
ignorance on the scale of the regularizing parameter $\alpha$. }
\label{priors}
\end{table}

In this work we assume flat spatial sections, and thus
$\Omega_m=1-\Omega_{DE}$. The parameters included in the
hypothesis space are summarized in Table \ref{priors}. These are
$\Omega_{DE}$, the Hubble parameter today, $H_0$ in km/s/Mpc, and
the coefficients of the DE decomposition, $C_i$, as described by
Eq.~\eqref{eq:coeff}. This generic characterization requires the
number of expansion functions (which can be effectively
characterized as top--hat bins) to be sufficiently large for this
to be a suitable description of $w(z)$. As described above, we
also include the nuisance parameters $\gamma = \log\alpha$ and the
value of the model EOS, $\model$, whenever this is marginalized
over.

Assuming uncorrelated Gaussian noise, the log--likelihood of a
point in parameter space is given by
 \be -2\log\ls (\wv)= \sum^{N_D}_{i=1}
\bra{\frac{D_i - f(z_i, \wv)}{\sigma_i}}^2,
 \ee
 where for each datum $i$ at redshift $z_i$
we have $f(z_i, \wv) \equiv H(z_i)$ for future radial baryonic
oscillation measurements, $f(z_i, \wv) \equiv  D_A(z_i)$ for
present and future transversal baryonic oscillation data and CMB
data and $f(z_i, \wv) \equiv  D_L(z_i)$ for SNIa data.
Furthermore, $\sigma^2_i$ is the measurement variance. The Hubble
parameter as a function of redshift is obtained via
Eq.~\eqref{Hz}, where the energy density is calculated using, for
$z_a$ falling within the $i$-th bin;
 \ber
\frac{\rho_{DE}(z_a)}{\rho_{DE}(0)}&=&\bra{\frac{1+z_a}{1+z_i-\Delta
z_i/2}}^{3(1+w_i)}  \nonumber \\
&\times& \Pi^{i-1}_{j=1}\bra{\frac{1+z_j+\Delta
z_j/2}{1+z_j-\Delta z_i/2}}^{3(w_j+1)}.
 \eer

From the above, the angular diameter distance $D_{A}(z_a)$ can
then be computed using equations (\ref{eq:1}-\ref{Hz}). For
piecewise constant $w(z)$ we employ the trapezoid rule to
approximate the integral, obtaining
 \ber
 D_{A}(z_a)&=&\frac{1}{1+z_a}
 \left[ \frac{\delta z}{4}
 \frac{1}{H_0} \right. \notag\\
 &+&\frac{1}{H(z_a)}+\sum^{a-1}_{j=2} \frac{\delta
z}{2}\bra{\frac{1}{H(z_{j-1})}+\frac{1}{H(z_j)}} \left.\right] .
\label{DA_2} \eer The binning of the integral, defined by $\delta
z$ is determined based on a fixed level of fractional accuracy for
the integration, that we set to $10^{-11}$ as determined by the
extrapolation error estimate. Finally, the entropy of a vector
$\wv$ with respect to a model $\model$ is given by
 \be \label{eq:entr}
 S(\wv) = \sum_{i=1}^{N} \bras{C_i - M_i - C_i \log \bra{\frac{C_i}{M_i}}}, \ee
 where the $M_i$, $i=1,\dots,N$ are the coefficient of the model
$\model$ in the expansion \eqref{eq:coeff}.

To sample the posterior probability distribution efficiently we
use a Monte Carlo Markov Chain (MCMC) which employs a Metropolis
algorithm. For more details about MCMC, see e.g.
\cite{Neal:1992jw, Lewis:2002ah}. Since the MaxEnt method is
designed to achieve the optimal reconstruction independently of
the number of degrees of freedom in the parameterization of
$w(z)$, we expect that the number of basis functions $N$ will not
affect greatly the reconstruction, as long as $N$ is chosen large
enough to capture the possible structure in the data. In the
following we choose $N=10$ but we have checked that the results do
not vary much if one uses $N=5$ or $N=20$ instead.

When using actual data, we divide the redshift range spanned by
either the SNLS or the HST/GOODS supernova measurements into
$N=10$ equally spaced bins, corresponding to the $N$ basis
functions for $w(z)$. We then extend the last bin to cover all of
the redshift range to last scattering when computing the angular
diameter distance to the CMB. In other words, we take $w(z)$ to be
constant (but not fixed to $-1$) between the redshift of the
highest supernova in the samples and $z = 1089$. This
extrapolation is weaker than the 'strong' prior used in the
analysis of \cite{Riess:2006fw}, which assumed that $w=-1$ at
$z>1.8$.

\section{Demonstration of the MaxEnt method}
\label{sec:demo}

We now proceed to test our MaxEnt reconstruction method with
synthetic data. Our benchmark dataset consists of $N_D=10$
measurements of $H(z)$ and $N_D=10$ of $D_A(z)$ (or equivalently,
$D_L(z)$) distributed uniformly in the redshift range $0 \leq z
\leq 1$. Although the actual measurements will in reality be less
homogeneous, this does not represent a problem for our
reconstruction algorithm, as we show below. Existing measurements
of $D_L(z)$ out to $z \sim 2$ will be vastly improved when future
surveys such as DES or LSST will be able to observe thousands of
SNIa per year \citep{Abbott:2005bi, Tyson:2006hs} and space--based
projects such as SNAP \citep{Aldering:2004ak}, ADEPT or DUNE will
provide observations at redshift beyond $z \sim 0.8$. Future
spectrographic surveys such as the Wide--Field----Multi--Object
Spectrograph (WFMOS) or HETDEX ought to deliver constraints on
$D_A(z)$ of $1\%$ at $z \sim 1$ and $1.5\%$ at $z\sim 3$
($1\sigma$) and $H(z\sim1)$ to $1.2\%$
\citep{Glazebrook:2005ui,Kelz:2006rs}, with better performance
still to be expected when the Square Kilometer Array will come
online \citep{Blake:2004pb}.

We add Gaussian noise to our synthetic data as a fixed percentage
of the true value of the measured quantity. We use an optimistic
noise level of $1\%$. Although it is beyond the scope of this
paper to make quantitative predictions about the performance of
future surveys in reconstructing the EOS, our benchmark data set
roughly reflects the potentiality of future observations. We also
test the performance of the method when the signal--to--noise
level is degraded by a factor of 10, in order to check for bias in
the reconstruction due to our entropic prior when the quality of
the data is poor. In this case we use a noise level of $10\%$ in
the luminosity distance and Hubble function measurements.
\begin{figure}
\centering
     \includegraphics[trim = 0mm 5mm 0mm 0mm, scale=0.7]{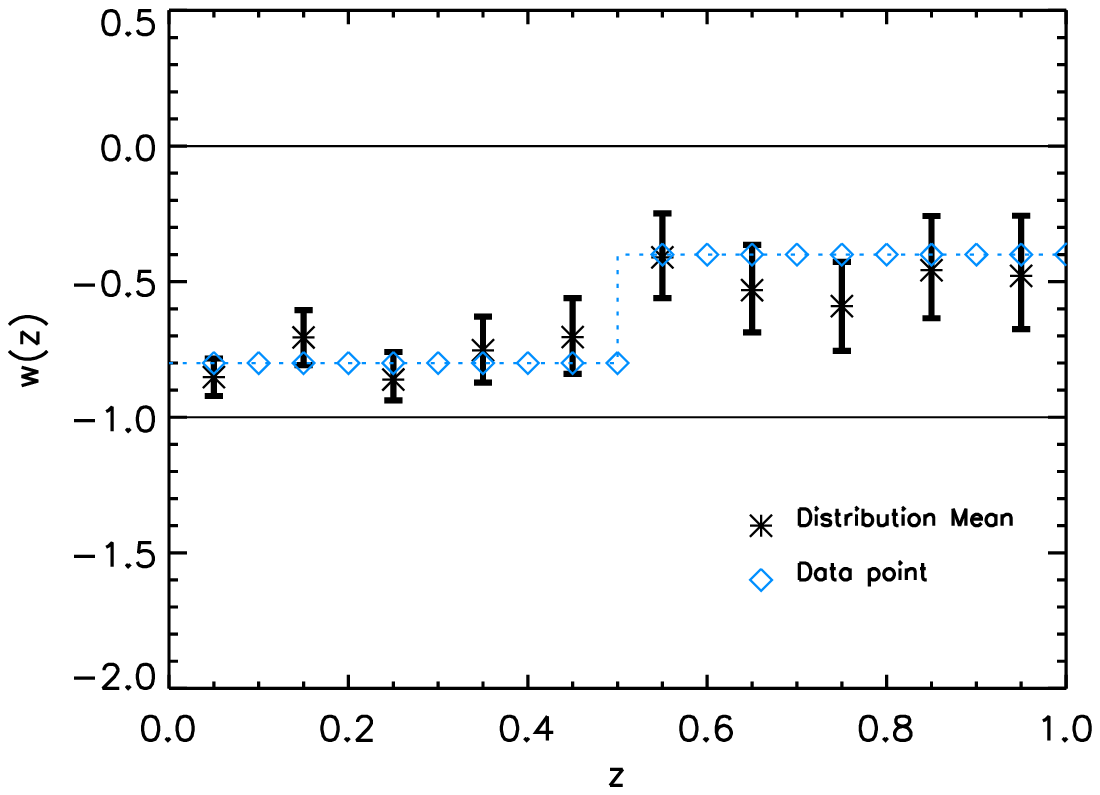}
     \includegraphics[trim = 0mm 5mm 0mm 0mm, scale=0.7]{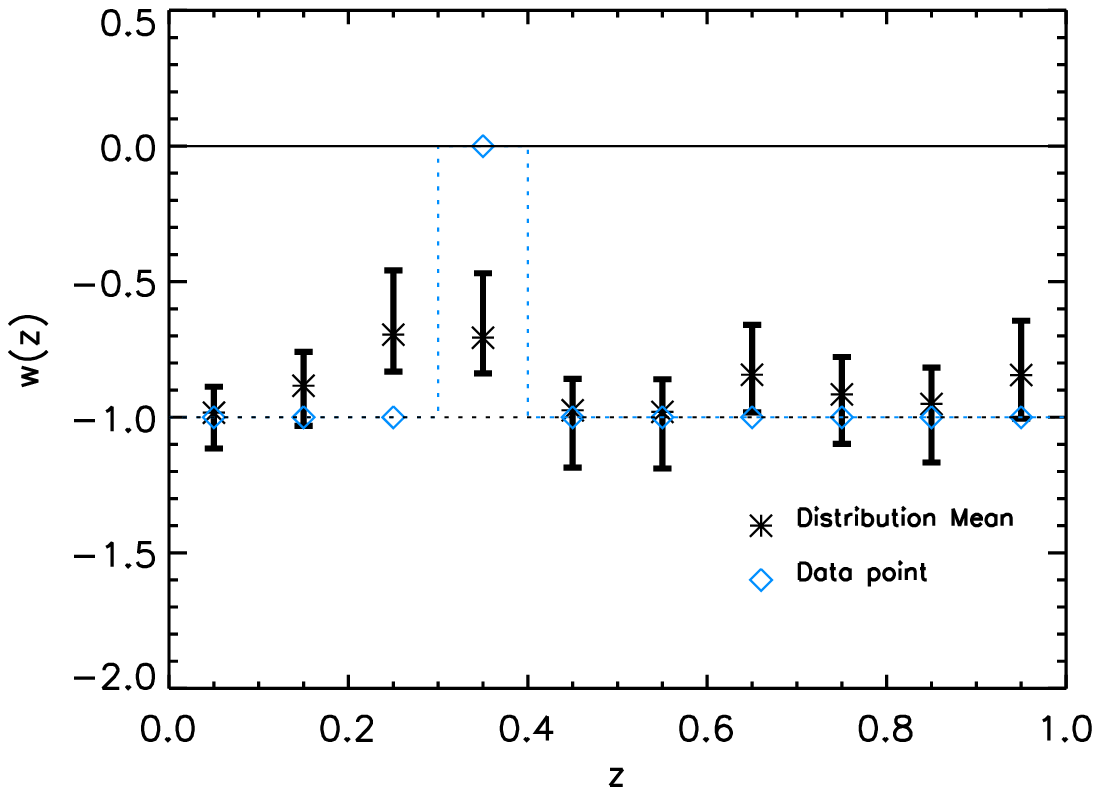}
     \includegraphics[trim = 0mm 5mm 0mm 0mm, scale=0.7]{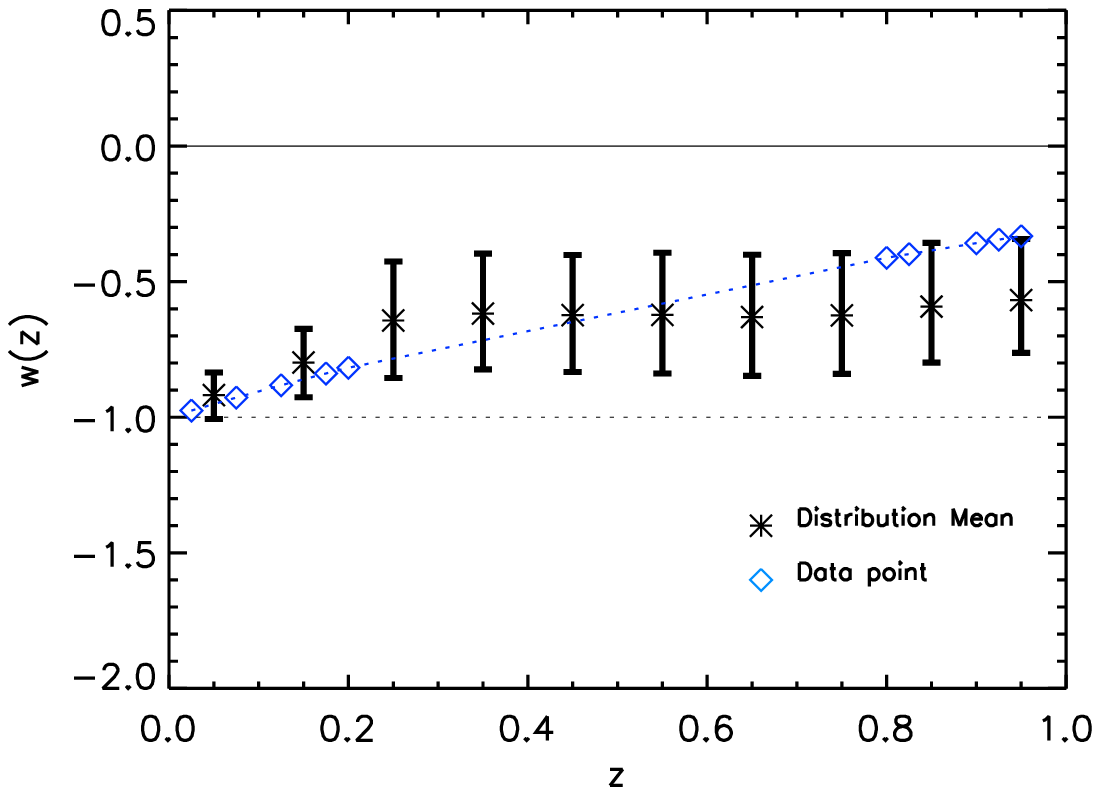}
\caption{Reconstructed EOS $\wv$ (black errorbars showing $1
\sigma$ posterior constraints) using our MaxEnt method for
high--quality synthetic data ($N_D=20$, $\sigma = 1\%$, location
shown by the blue diamonds) in the redshift range $0 \leq z \leq
1$. Top panel: the true EOS (blue, dotted line) is taken to be a
step function. Middle panel: the true EOS shows a sharp peak at $z
\sim 0.4$.  Lower panel: the true EOS is slowly evolving with
redshift, and the synthetic data are now clustered at low and high
redshift. Despite the absence of data points at intermediate
redshifts, the high-$z$ reconstruction tracks the true function
with reasonable accuracy, while the intermediate redshift errors
increase correspondingly. In all three cases, the value of the
prior model $\model$ for $w(z)$ has been marginalized over and the
MaxEnt reconstruction satisfactorily recovers the true EOS. We
have plotted horizontal lines at $w=0$ and $w=-1$ to guide the
eye.} \label{fig:highquality}
\end{figure}

We show in Figure \ref{fig:highquality} the reconstructed EOS for
our benchmark scenario with high quality observations ($\sigma =
1\%$, $N_D=20$ observations). In all three panels, we have
marginalized over the prior model $\model$. We notice that the
reconstruction is satisfactory in all three cases. We have checked
that the $\wv$ extracted when marginalizing over the prior model
$\model$ has comparable accuracy to the case of a fixed model
$\model$. The bottom panel shows how the method deals with gaps in
the redshift range of the observations: the smoothing effect of
the entropic term enlarges the errors in the region where no data
are available, while the reconstructed EOS tracks the true value
at small and large redshifts, where the data are clustered. Here
we have employed a 10--dimensional $\wv$, but we have checked that
increasing the number of elements to $N=20$ does not lead to any
significant change in the reconstruction. As expected, the error
bars in the regions where $w(z)$ is well constrained by
observations are considerably smaller.
\begin{figure}
\centering
     \includegraphics[trim = 0mm 5mm 0mm 0mm, scale=0.7]{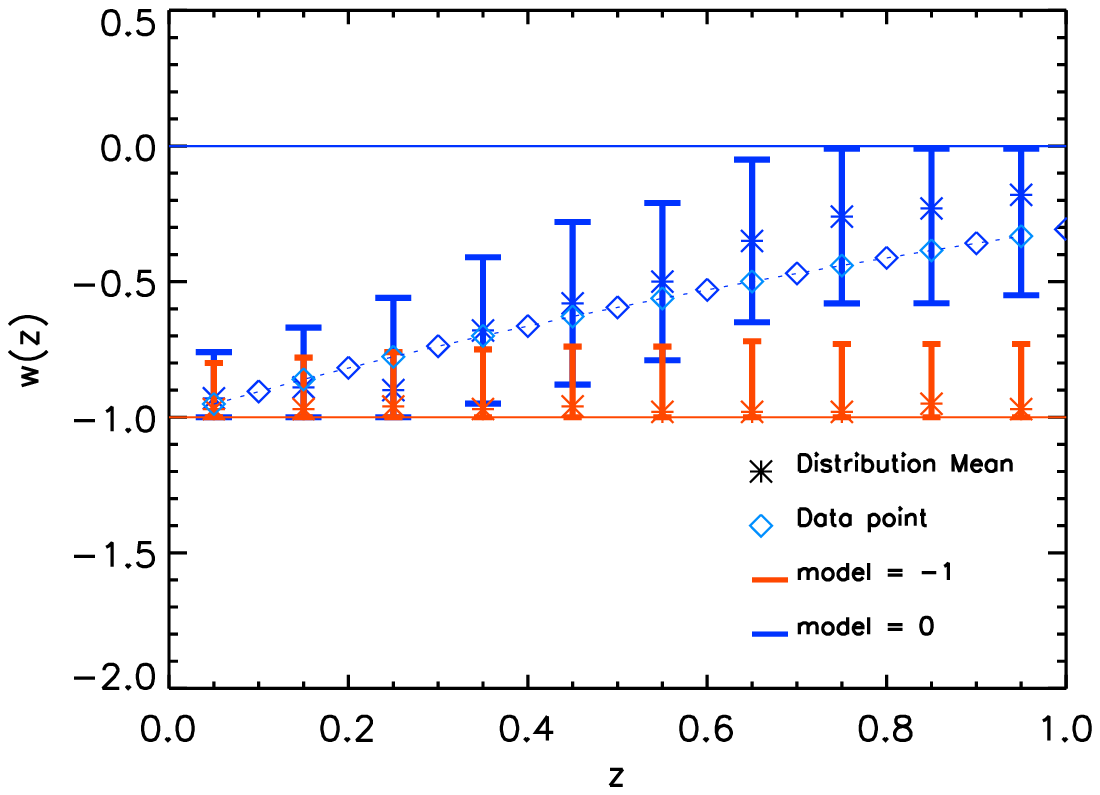}
     \includegraphics[trim = 0mm 5mm 0mm 0mm, scale=0.7]{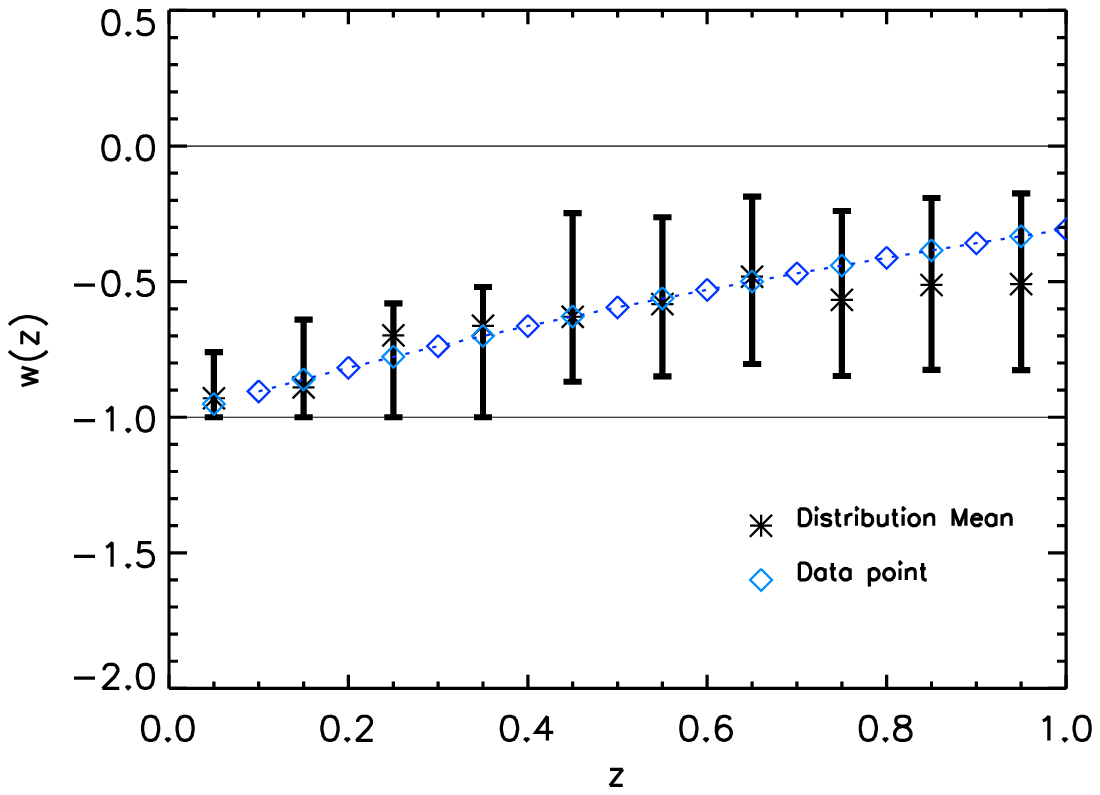}
\caption{Reconstructed EOS $\wv$ (errorbars showing $1 \sigma$
posterior constraints) for noisy synthetic data ($N_D=20$, $\sigma
= 10\%$, location shown by the blue diamonds). Top panel: the
reconstructed EOS for a model $\model = -1$ (red errorbars) has
collapsed towards the model due to the entropic prior for
redshifts $z \gsim 0.3$, while the reconstruction with $\model =
0$ (blue errorbars) tracks better the true EOS (blue, dotted
line). Bottom panel: after marginalization over the model value,
the bias in the reconstruction has disappeared, but the errorbars
have become suitably larger.} \label{fig:lowquality}
\end{figure}

We now turn to the case where the data are noisy, and hence we
expect our entropic prior to play a more important role in the
reconstruction. The top panel of Figure \ref{fig:lowquality} shows
the reconstructed EOS for a noisy data set of $N_D = 20$
measurements with noise $\sigma = 10\%$ and a slowly evolving true
EOS, $\wv(z) = 1 - \ln(1+z)$. In the top panel, we show the result
when employing as prior model values the two liming cases of $w$
with strong theoretical prejudice: $\model=-1$ (cosmological
constant, red errorbars) and $\model=0$ (Einstein-de Sitter
Universe, blue errorbars). In the high $z$ bins the reconstruction
becomes increasingly mismatched with the underlying true EOS.
Because the dependence of the data on the EOS requires integrating
the EOS over redshift, any error in the reconstruction at low
redshift is accumulated as $z$ increases. As a result the entropy
tends to dominate over the likelihood and the mean parameter
values collapses towards the model at higher redshift, especially
for the case where $\model = -1$. Even though the reconstruction
has appeared to degrade for the $\model = 0$ case, it is
encouraging that the mean values of the parameters in the lower
redshift bins ($z \lsim 0.3$) are reasonably close to the true
values.

Evidently the choice of prior model $\model$ does have some
bearing on the reconstructed value of the parameters at high
redshift (given that the entropy dominates the posterior for
poorly informative data) and this must be kept in mind when
interpreting the results when dealing with noisy data. This
example highlights the problem of distinguishing a genuine
affinity for a certain function that happens to closely resemble
the model from a strong default towards the model on account of
noisy observations, i.e.~how will we interpret a result very close
to $w(z)=-1$? In this case, there are two options: the
observational evidence is noisy and uninformative, leading to
entropy domination, or the data is good and favours an actual
value close to $-1$. In the latter case, attempting a
reconstruction with $\model = 0$ will allow to test the strength
of the data in pulling $w(z)$ towards the cosmological constant
value. An alternative mean of recovering $w(z)$ in a truly
model--independent way is to include the elements of the model
vector in the hypothesis space, as discussed in section
\ref{sec:mod}. Marginalizing over $\model$ then amounts to
selecting the optimum distribution at each sampled point. The
model still represents the most entropic distribution given that
it is uniform in $z$--space. The result of this procedure is shown
in the bottom panel of Figure~\ref{fig:lowquality}, where
marginalization over the model has cured the skew in the
reconstruction observed above, albeit at the price of delivering
larger error bars.

\section{Application to present-day data}
\label{sec:application}

We now apply our reconstruction procedure to actual data,
encompassing cosmic microwave background observations, baryonic
acoustic oscillations measurements, determination of the present
value of the Hubble parameter and two different supernovae type Ia
samples.

\subsection{Data sets}

\subsubsection{Baryon Acoustic Oscillations}

The comoving sizes along the line of sight, $r_{||}$ (in redshift
space), and in the traverse directions, $r_{\perp}$, of a feature
sitting at a redshift $z$ are related to the redshift range
$\Delta z$ covered and the angle subtended $\Delta \Theta$,
respectively, by
  \be
 r_{||}=\frac{c \Delta
z }{H(z)}, \quad r_{\perp}=(1+z)D_{A}(z) \Delta \Theta.
 \ee
 If the absolute values of
$r_{\perp}$ and $r_{||}$ are known, they become standard rulers
giving us a handle on $H(z)$ and $D_{A}(z)$. If only the relative
sizes are known, then the standard rulers are expressed in units
of $H_0$. If only the ratio $r_{||}/r_{\perp}$ is known, this
becomes the Alcock--Paczynski test. The baryon acoustic
oscillation phenomenon can be used as such a standard ruler.

After recombination, when the Universe becomes neutral and photons
free--stream from the cosmic plasma, the driving force of the
harmonic oscillation is removed and the sound speed of the
now-neutral medium essentially falls off to zero, ending wave
propagation. The spatial distribution of the baryons at this stage
will then reflect the characteristic scale of the acoustic waves.
Seeing as the perturbations in the baryon and CDM distributions
seed the formation of large scale structure, we expect to see
acoustic peaks in the late-time matter power spectrum
\citep{Eisenstein:2005su}. This becomes a standard ruler because
the scale of these acoustic oscillations is self-calibrated under
standard recombination \citep{Hu:2004kn}. It depends solely on the
photon--baryon ratio and radiation--matter ratio at recombination
which are determined with excellent precision in the CMB power
spectrum  from the CMB peak morphology \citep{Eisenstein:2004an}.
The change in the apparent size of this scale from recombination
to the present will depend on the expansion history of the
universe through projection effects. These acoustic features
appear as rings in angular and redshift space \citep{Hu:2003ti}.
The actual measurement from the SDSS LRG sample is of the dilation
factor, defined as
 \be D_V(z)=\bras{D^2_A(z) \frac{cz}{H(z)}
}^{1/3} \label{Dv} \ee where the comoving angular diameter
distance $D_A$ is taken as the transverse dilation and the line of
sight measurement of this scale $\frac{cz}{H(z)}$ is taken to be
the radial dilation. The observed correlation scale constrains a
function of the dilation factor, and a single data point is
reported in \cite{Eisenstein:2005su}: \be A \equiv D_V (0.35)
\frac{\sqrt{\Omega_m H^2_0}}{0.35 c}=0.469 \pm 0.017. \ee (see
also \cite{Cole:2005sx} for a similar detection of the acoustic
feature in the 2dF catalogue). The above assumes $\Omega_b
h^2=0.027$. The log--likelihood for the baryon acoustic
oscillation data is given by \be
\chi^2_{BAO}=\frac{(A-0.469)^2}{0.017^2}. \ee

\subsubsection{The Supernova (SNIa) data}

Type Ia supernovae are good candidates for standard candles and
are useful in determining distances on extragalactic scales. Due
to the complexity of the physics involved, the SNIa are not
perfect standard candles, having a dispersion of $0.3 - 0.5$ mag
in their peak magnitudes \citep{Straumann:2006tv}. However the
peak brightnesses appear to be tightly correlated to the
time-scale of their brightening and fading and one can extract an
empirical relation between absolute peak luminosity and the
morphology of the light curves to constrain the absolute
brightnesses, and thus obtain measurements of $D_L(z)$.

The first group of supernovae, termed the `gold' set, is from the
HST/GOODS programme \citep{Riess:2004nr}, complemented by the
recently discovered higher redshift supernovae, reported in
\cite{Riess:2006fw}, while the second sample is taken from the
Supernovae Legacy Survey (SNLS) \citep{Astier:2005qq}. As
discussed in e.g.~\cite{Wang:2006ts}, it appears that there are
systematic differences between these two data sets that arise from
differences in the data processing. It is therefore necessary to
consider the two datasets separately, and compare the results as a
consistency check.
\\
\emph{The 'gold' sample:}\\
The distance modulus $\mu$ is defined as the difference between
the apparent magnitude $m$ and the absolute magnitude $M$:
 \be \mu = m - M= 5 \log_{10} \bra{\frac{D_{L}(z)}{10 \text{pc}}}.
  \ee
Given that the absolute magnitude $M$ is a unknown, we consider
the following distance modulus $\mu_i(\delta M)$:
 \be
\mu_{i}(\delta M)=\mu^d_{i}-\delta M. \ee Here $\delta M$ is the
difference between the mean of the true absolute magnitudes and
the estimated absolute magnitude, while the $\mu_{i}^d$ are the
observed magnitudes after dust corrections and recalibration
through the shape of the luminosity evolution function. The
quantity $\delta M$ is the difference between the true mean
absolute magnitude and the estimated absolute magnitude of the
supernovae and is marginalized over.
\\
\emph{The SNLS sample:}\\
The SNIa data from the Supernovae Legacy Survey (SNLS) are reduced
in a different manner in that the light curves provide constraints
on various parameters which are then used to calculate the
effective apparent magnitude. For a description of the calculation
of $\mu_i$ see \cite{Dick:2006ev}. From the observed $\mu_i$ with
variances $\sigma^2_{\mu,i}$  for each set of SNIa, we perform an
analytical marginalization over the absolute magnitude $M$.
Defining the quantities \be f_0 = \sum_{i=1}^{N} \frac{5 \log_{10}
D_L - \mu_i}{\sigma^2_{\mu,i}}, \ee \be c= \sum_{i=1}^{N}
\frac{1}{\sigma^2_{\mu,i}} \ee and \be f_1=\sum_{i=1}^{N} \frac{(5
\log_{10} D_L(z_i) - \mu_i)^2}{\sigma^2_{\mu,i}} . \ee the
M-independent log-likelihood for the SNIa is calculated as \be
\chi^2_{SN}=f_1-\frac{f_0^{2}}{c}. \ee

\subsubsection{CMB and HST data}

The WMAP 3-year measurement of the CMB shift parameter describing
the location of peaks in the CMB power spectrum serves to
constrain the angular diameter distance to last scattering
\citep{Spergel:2006hy}. This is independent of most assumptions of
the form of dark energy, and is given by \citep{Wang:2006ts} \be
R=\Omega_m^{1/2}H_0 \int^{z_{CMB}}_{0}\frac{dz}{H(z)} = 1.70 \pm
0.03, \ee where $z_{CMB}$ is the redshift to last scattering,
taken in our case to be $1089$.

We also include the constraint on the present value of the Hubble
constant obtained  by the Hubble Space Telescope Key Project
\cite{Freedman:2000cf}, by using a Gaussian likelihood with mean
and standard deviation given by $H_0= 72 \pm 8$ km Mpc s$^{-1}$.

\subsection{Results}

\begin{figure}
\centering
  \includegraphics[trim = 0mm 5mm 0mm 0mm,scale = 0.7]{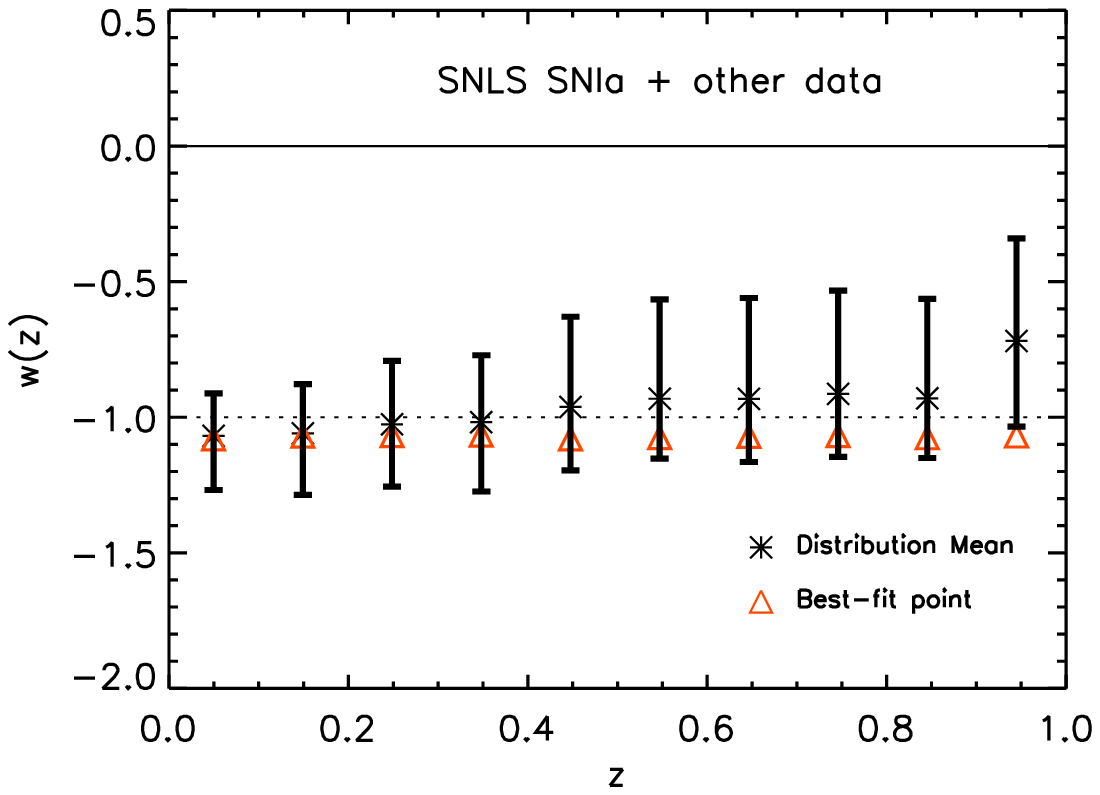} \\
  \includegraphics[trim = 0mm 5mm 0mm 0mm,scale = 0.7]{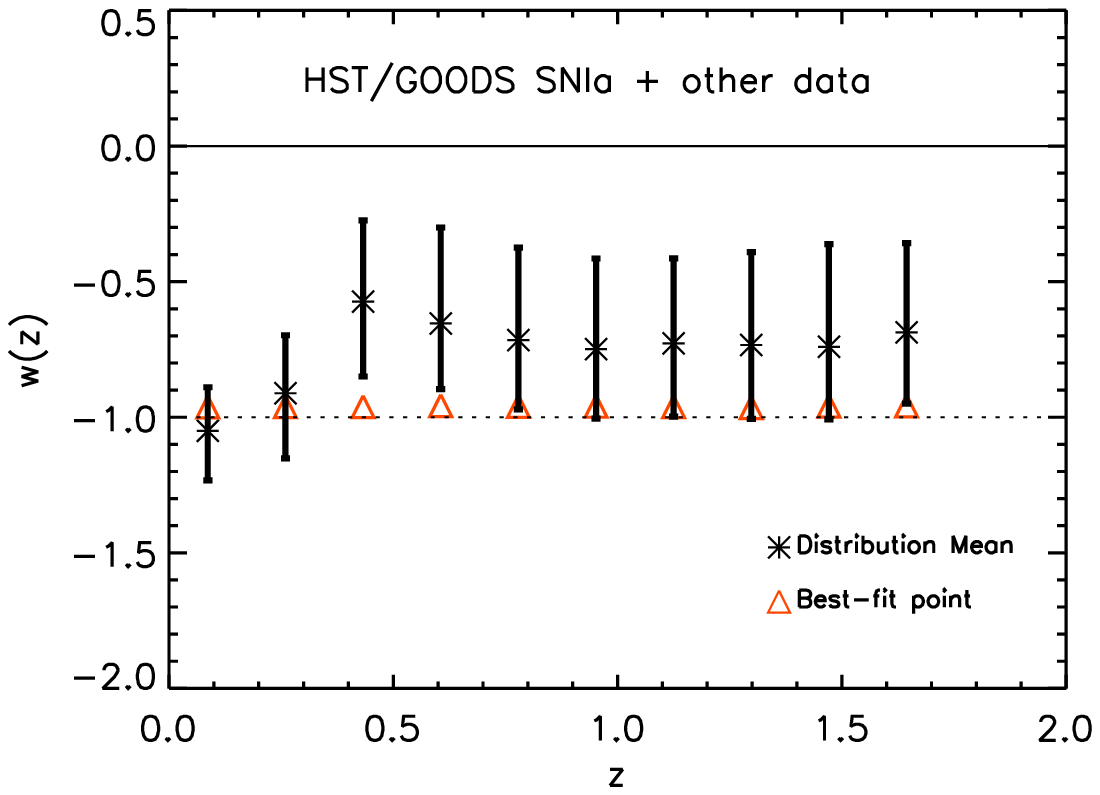} \\
\caption{Reconstructed EOS (marginalized errorbars encompassing
$68\%$ of posterior probability) from SNLS data (top panel) and
HST/GOODS data (bottom panel), including CMB, HST and BAO
measurements, as well (notice that the redshift range is different
for the two panels). The prior model $\model$ is constant in
redshift and has been marginalized and the assumed range of the
EOS is $-2 \leq \wv \leq 0$ (both are conservative choices). The
horizontal lines indicate the upper bounds of the allowed $\wv$
range (solid) and position of $w=-1$ (dotted)  in order to guide
the eye. The SNLS data do not show significant deviations from $w
= -1$ over the whole redshift range.  The reconstruction from the
HST/GOODS data is also consistent with a cosmological constant,
although it appears to slightly prefer a higher value at redshift
$z \sim 0.5$. The best-fitting points are all very close to
$w=-1$.} \label{fig:snls_vs_goods_1}
\end{figure}

\begin{figure}
\centering
  \includegraphics[trim = 0mm 5mm 0mm 0mm,scale = 0.4]{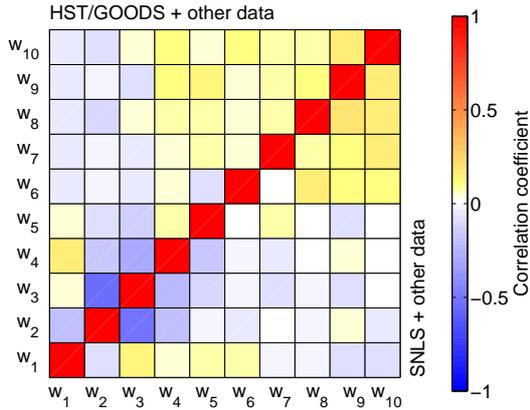}
\caption{Correlation matrix for the reconstructed EOS parameters
from HST/GOODS (upper left) and SNLS (lower right) data sets,
including CMB, HST and BAO measurements. The strongest
(anti)correlations are between values in bins 2 and 3 which
roughly coincide with the redshift position of the BAO
measurement.} \label{fig:covariance} \label{fig:cov_matrix}
\end{figure}

We plot in Figure~\ref{fig:snls_vs_goods_1} the results of our
reconstruction from the SN type Ia data from the SNLS and from the
HST/GOODS programmes. In both cases we have added the CMB, HST and
BAO measurements, and we have marginalized over the model, in
order to be as conservative as possible. Furthermore, the maximum
range for the EOS has been taken to be $-2 \leq \wv \leq 0$. We
plot regions encompassing $68\%$ of posterior probability for each
$\wv$ bin -- notice that since these are marginalized values,
their magnitude is independent of the correlations between
reconstructed points (the issue of correlations is addressed in
detail below, see Figure~\ref{fig:cov_matrix}).

We do not find any significant deviation from a cosmological
constant behaviour from the SNLS data (see top panel of
Figure~\ref{fig:snls_vs_goods_1}) at all redshifts. Our posterior
constraints in this case are rather conservative, as a consequence
of the assumptions made in the reconstruction (i.e., large $\wv$
range and marginalization over the model).  When using the
HST/GOODS data, the recovered $\wv$ in the first bin agrees with
that found for the SNLS data. Given that a number of the SNIa in
this bin are common to both surveys, this provides a consistency
check. The reconstructed EOS from the HST/GOODS data is however
found to prefer slightly higher values in the third bin ($z \simeq
0.5$), excluding the cosmological constant value to a little bit
more than $1\sigma$ significance. The significance of this rise
however has to be assessed with care, especially if we recall that
the constraining power of the SNIa data degrades in this region
\citep{Simpson:2006bd}. At redshifts above $z\sim 0.5$, the mean
of the reconstruction settles around $w \sim -0.7$, although we
notice that the best fitting points remain very close to $w=-1$
(red triangles in Figure~\ref{fig:snls_vs_goods_1}). The implied
early-time behaviour of dark energy is consistent with the result
of $w=-0.8^{+0.6}_{- 1.0}$ found in \cite{Riess:2006fw} for $z>1$,
using what they call their 'strong' prior. If instead the EOS is
assumed to be constant over the entire redshift space (i.e., if we
reduce the number of $w$ components to $N=1$), then we obtain from
the HST/GOODS data $w = -0.89 \pm 0.07$, in agreement with usual
results (see e.g. \cite{Riess:2006fw}). This clearly demonstrates
the danger of assuming $w(z)$ to be time--independent, as one
would miss in this way possible features in the data.

The use of the entropic prior introduces correlations among the
reconstructed points (see \cite{Huterer:2004ch} for a technique to
extract uncorrelated band power estimates of the EOS). The
correlation coefficients from the posterior distribution over the
$\wv$ parameters are shown for both data sets in
Figure~\ref{fig:cov_matrix}. We notice that correlations are in
general relatively mild, flattening around the level of $\sim
20\%$ for correlations with bins at larger redshifts, where the
entropic prior becomes more important. The strongest correlations
(at the level of $\sim 50\%$) are observed among parameters in the
2nd and 3rd redshift bins, where the BAO measurement strongly
constrains the EOS and due to the fact that the observables are
integrated over redshift, we expect a negative correlation among
the well--constrained value at the position of the BAO measurement
and the $\wv$ values at lower redshift.

\begin{figure}
\centering
  \includegraphics[trim = 0mm 5mm 0mm 0mm,scale = 0.7]{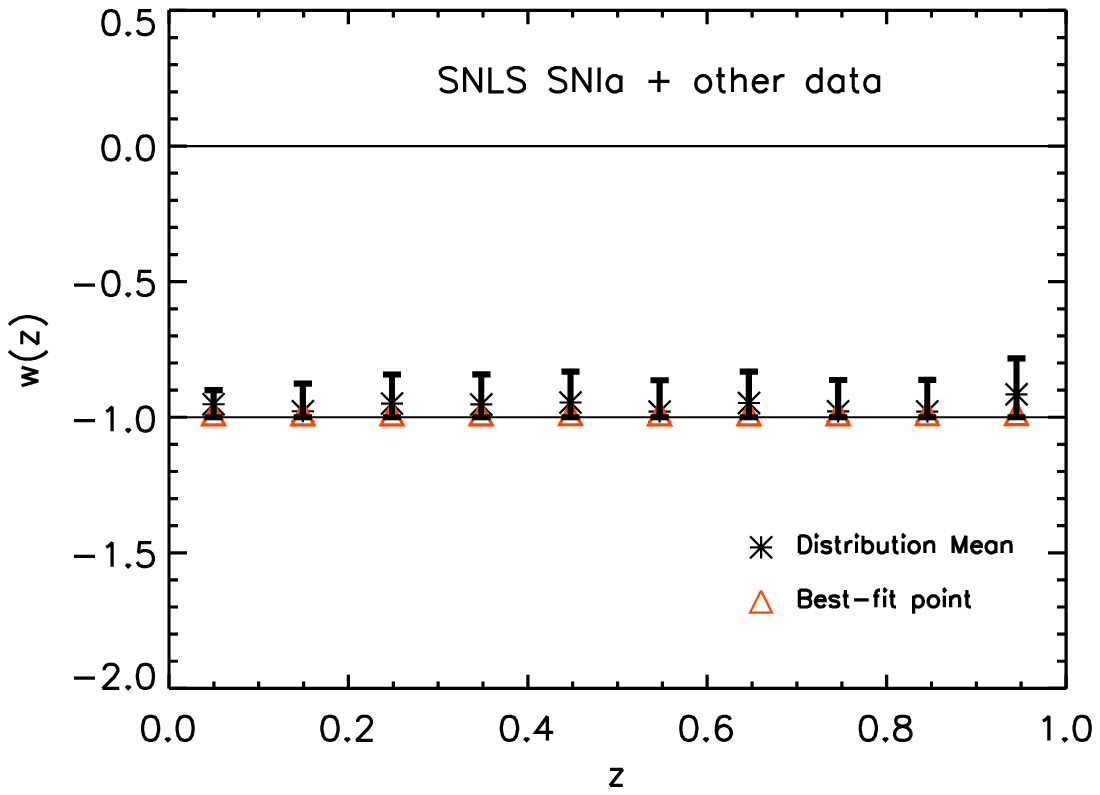} \\
  \includegraphics[trim = 0mm 5mm 0mm 0mm,scale = 0.7]{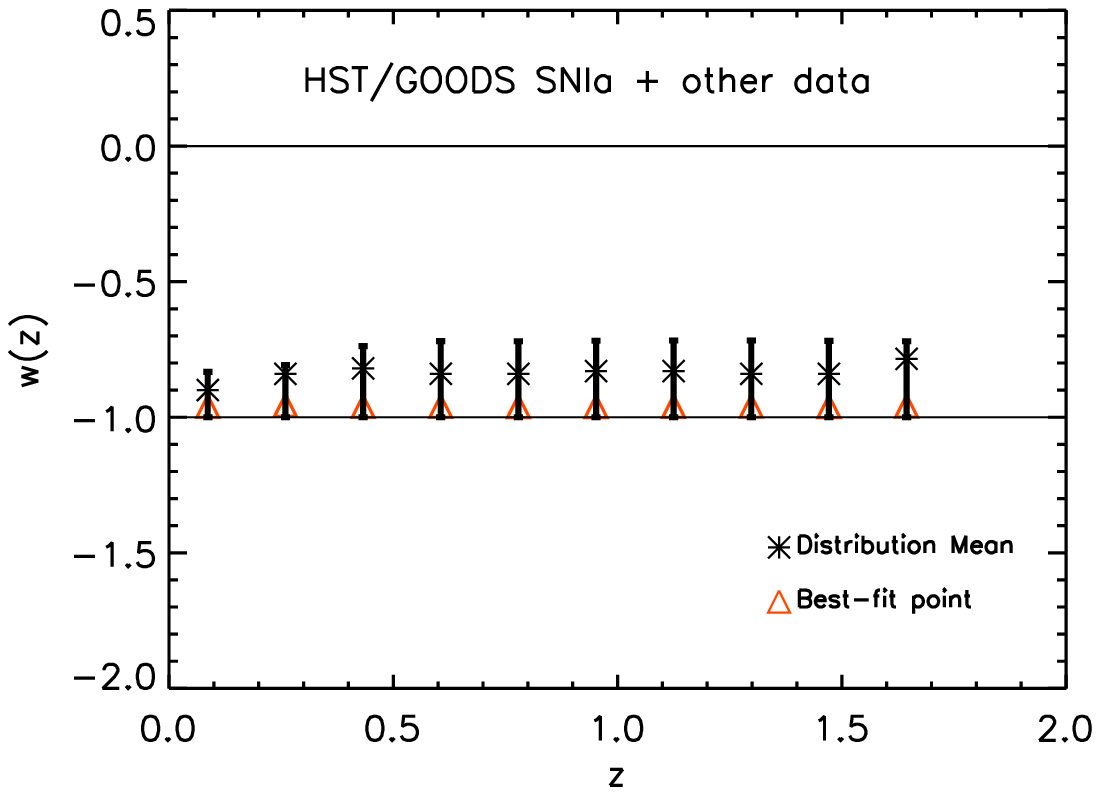} \\
\caption{Reconstructed EOS (marginalized errorbars encompassing
$68\%$ of posterior probability) from SNLS data (top panel) and
HST/GOODS data (bottom panel), marginalizing over a constant prior
model but restricting the $w$ range to $-1 \leq \wv \leq 0$. The
horizontal lines indicate the upper and lower bounds of the
allowed $\wv$ range (solid) in order to guide the eye. We find no
significant deviation from $w=-1$ for the SNLS data set. For the
HST/GOODS sample the reconstruction tends to drift to larger
values at higher redshift.} \label{fig:snls_vs_goods_2}
\end{figure}

We now investigate the case where we impose that $w\geq -1$ on our
parameter space. The results are shown in
Figure~\ref{fig:snls_vs_goods_2}, where the reconstructed EOS
using the SNLS tracks the cosmological constant value at all
redshifts, with $1-$tail $1\sigma$ errors of order $0.2$ at all
$z$ values. Because of the reduced freedom in $\wv$, the
reconstruction collapses to the lower limit of the allowed $w$
range, even after marginalization over the model values. In the
case of the HST/GOODS data, a gentle rise of $w(z)$ away from $-1$
is again observed. The larger error bars suggest that the entropy
becomes important and that the value  $\wv\sim -0.8$ to which the
reconstruction tends at redshifts $z\gsim 0.5$ is mediated by the
the mean value of the prior model $\model$. As before, the best
fit points remain very close to $w=-1$ at all redshifts.

\begin{figure}
\centering
  \includegraphics[trim = 0mm 5mm 0mm 0mm,scale = 0.7]{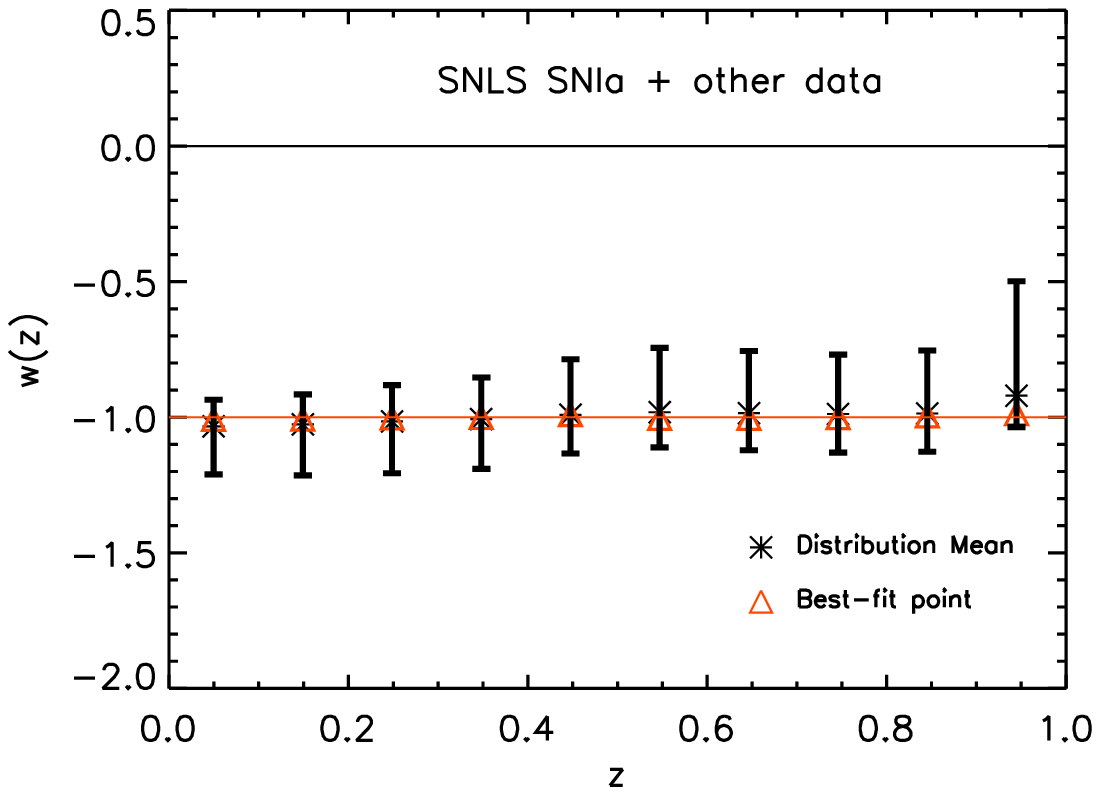} \\
  \includegraphics[trim = 0mm 5mm 0mm 0mm,scale = 0.7]{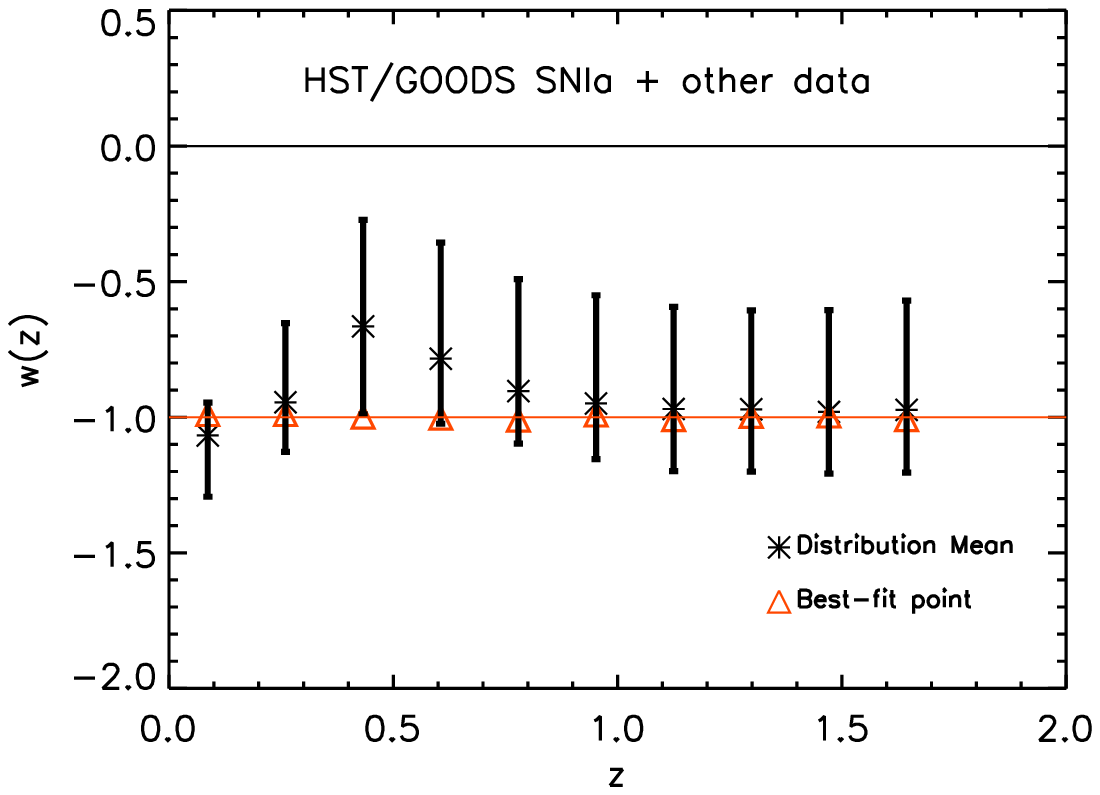} \\
\caption{Reconstructed EOS from SNLS data (top panel) and
HST/GOODS data (bottom panel), assuming an entropic prior model
$\model =-1$ and a $w$ range $-2 \leq \wv \leq 0$.  The horizontal
lines indicate the upper bound of the allowed $\wv$ range (black)
and the model $\model$ (red). The SNLS data are compatible with
the model and show errorbars of order $20\%$ at all redshifts. The
slight bump at $z\sim 0.5$ for the HST/GOODS data survives the
entropic prior.} \label{fig:snls_vs_goods_3}
\end{figure}

We can increase the amount of prior information by considering the
case where a constant prior model value $\model =-1$ is used, see
Figure~\ref{fig:snls_vs_goods_3}. This is helpful in assessing
whether the structure observed in the HST/GOODS sample is strong
enough to override our entropic prior. The reconstruction from the
SNLS data remains close to  $\wv = -1$ with errorbars of order
$20\%$ at all redshift. One has however to keep in mind that the
tightness of the errors is partially helped by the supplementary
information provided by the entropic prior. This demonstrates how
the use of cosmological constant as the model can be problematic
as one can not say conclusively whether this indicates that the
data is very strong or alternatively overridden by the entropy if
no significant deviations from the model are observed. The result
from the HST/GOODS data shows again a high value of $w(z)$ being
favoured in the third bin and subsequent collapse towards to the
model $\model$ at higher redshifts. The persistence of a deviation
towards $\wv>-1$ at redshift $z \sim 0.5$ in the presence of the
strong prior favouring the cosmological constant suggests it is a
real feature of the data.

\begin{figure}
\centering
      \includegraphics[trim = 0mm 5mm 0mm 0mm,scale = 0.7]{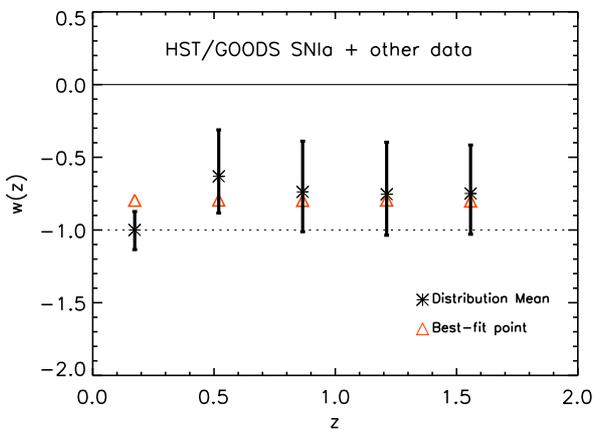} \\
\caption{Reconstructed EOS from the HST/GOODS data with 5
components in the reconstructed $\wv$. Compare with the bottom
panel of Figure~\ref{fig:snls_vs_goods_1}.}
\label{fig:snls_5_bins}
\end{figure}

Finally, we also investigated the stability of the reconstruction
against a change in the number of reconstructed components. Since
the MaxEnt technique is designed to automatically deal with the
structure in the data by adjusting the degree of smoothness of the
reconstruction, we do not expect that a change in the number of
bins would make a large difference in the reconstructed EOS. This
is demonstrated in Figure~\ref{fig:snls_5_bins}, where the
reconstruction analogous to the bottom panel of
Figure~\ref{fig:snls_vs_goods_1} has been performed by halving the
number of bins to $N=5$, without appreciable differences in the
end result.

Lastly, we investigate the dependency of our results on the model
one chooses for the entropic prior. Up until this point we have
assumed the prior model $\model$ to be constant with redshift such
that it introduces a suitable degree of smoothing of
time--dependent noisy features in the data.  This choice reflects
a specific belief in the true form of the EOS and given the large
range of dark energy models on the market, it is important to
assess the impact of a different prior model $\model$. Another
popular class of models is given by an equation of state that is a
smooth varying function of redshift, such as
 \be m(z) = w_0 +
 w_1 \frac{z}{(1+z)}. \ee
Here the assumption is that the true EOS is a function of time and
evolves sufficiently slowly such that it may be effectively
characterized in a phenomenological way by the two parameters
$w_0$ and $w_1$~\citep{Sahni:2006pa}. This particular function is
a good approximation to many dark energy models but clearly it is
limited to how well it can cope with a rapidly evolving EOS
\citep{Liddle:2006kn}. Following e.g.~\cite{Ichikawa:2005nb}, we
impose the further constraint that the early Universe is
matter--dominated, i.e. we impose the condition $w_0+w_1< 0$ on
the prior model. In order to be as conservative as possible, we
again marginalize over both prior model parameters, $w_0$ and
$w_1$, as follows. The expression for the entropy is again given
by Eq.~\eqref{eq:entr}, but the model coefficients are now given
by
 \be C_i = (C^{m}_0 -
2) + (C^{m}_1 - 2) \frac{z_i}{(1+z_i)} +2 .
 \ee
 Here $C^{m}_0$ and $C^{m}_1$ are the coefficient representing
the prior model parameters. They are included in the hypothesis
space, then marginalized over.

\begin{figure}
\centering
      \includegraphics[trim = 0mm 5mm 0mm 0mm,scale = 0.7]{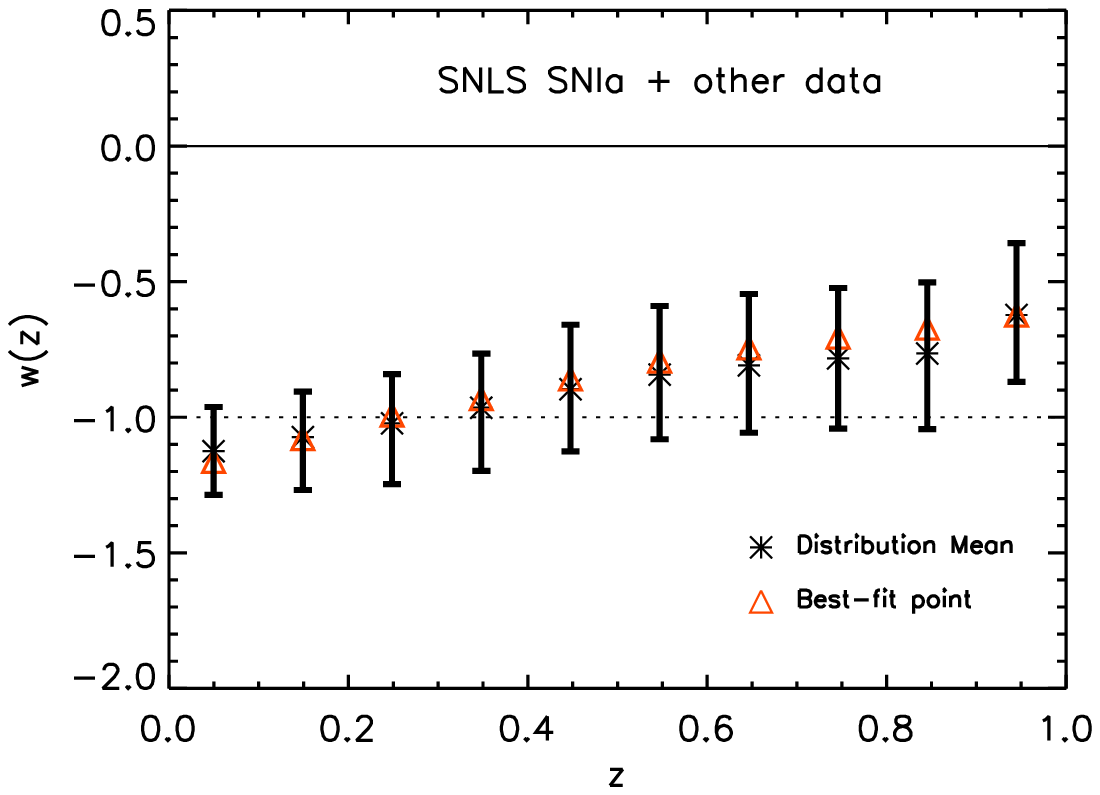} \\
      \includegraphics[trim = 0mm 5mm 0mm 0mm,scale = 0.7]{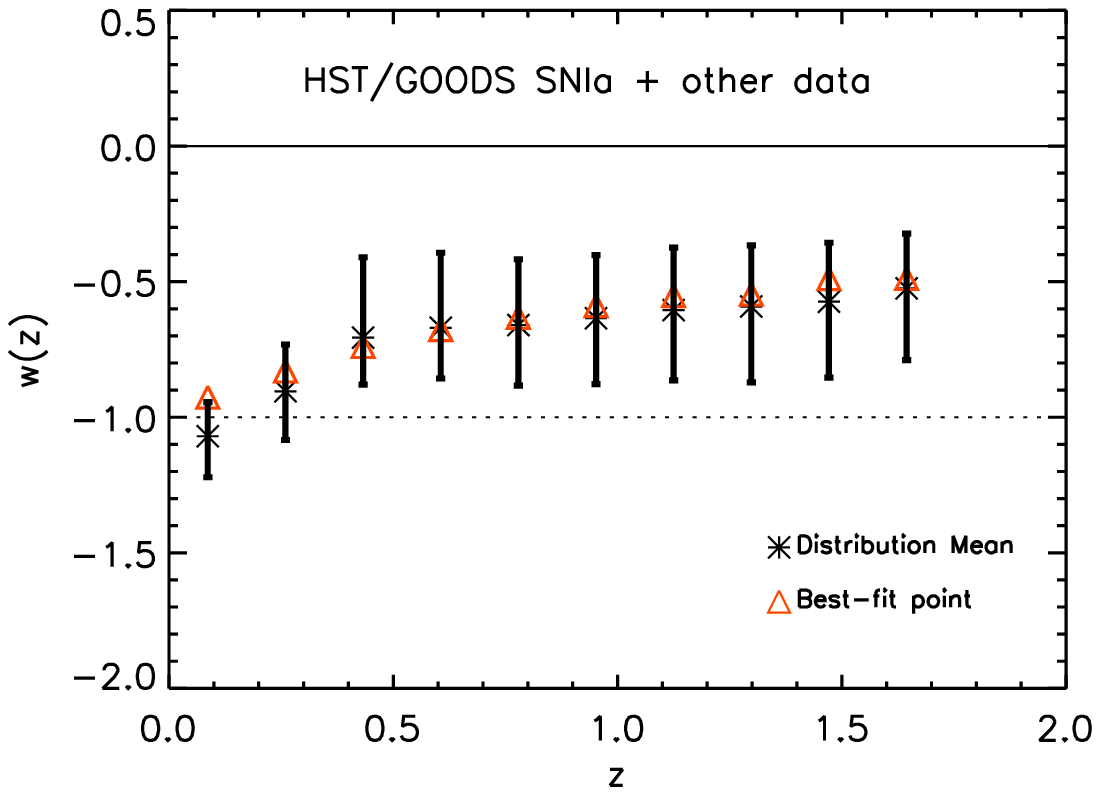} \\
\caption{Reconstructed EOS (marginalized errorbars encompassing
$68\%$ of posterior probability) from the SNLS data (top panel)
and the HST/GOODS data (bottom panel) marginalizing over a prior
model of the type $(w_0, w_1)$. The horizontal lines indicate the
upper bound of the allowed $\wv$ range (solid) and position of
$w=-1$ (dotted) in order to guide the eye. Comparison with
Figure~\ref{fig:snls_vs_goods_1}, where a constant prior model was
used, indicates that the results from both data sets are robust
against the choice of the prior model while exhibiting a mild
dependence on this choice at higher redshift, where the
constraining power of the data degrades and entropic domination
sets in. } \label{fig:model_w0w0_2}
\end{figure}

We display our results using this slowly--evolving prior model in
Figure~ \ref{fig:model_w0w0_2} (with the model parameters
marginalized over). This is to be contrasted with the analogous
case of Figure~\ref{fig:snls_vs_goods_1}, where the entropic model
$\model$ is constant in redshift space. At low redshifts the
reconstructions for different prior models do not differ
appreciably, with the most notable difference being a shift
towards slightly higher values of $w$ for the case of the SNLS
data set. For the GOODS/HST programme data set, the peak in the
third bin is recovered but is lower than in the previous constant
$\model$ case. Again, the reconstructed EOS prefers higher values
in the last few bins. At high redshifts, due to the accumulation
of error, we expect more entropic domination and hence a stronger
weight of the prior model choice. However in this region dark
energy becomes progressively less important and the ability of
data to constrain the time--depedence of the EOS thus degrades
considerably. It is encouraging however that the low redshift
behaviour of the reconstructed EOS agrees well for both
marginalized prior models. This further strengthens our
conclusions regarding the power and robustness of our technique.

\section{Conclusions}

\label{sec:conclusions} Given that the dark energy models on the
market are predominantly phenomenological, a reconstruction
technique that does not require specifying a parametric form for
$w(z)$ would be an advantage. We presented such a method based on
maximum entropy to reconstruct the equation of state of dark
energy within a Bayesian framework. The principle of maximum
entropy is invoked when assigning the Bayesian prior. This means
that in the absence of genuine signal, the model $w(z)$ that is
most smooth, or that maximizes the information entropy is
favoured, with the extent of this bias being determined by a
regularizing constant that is automatically adjusted to the data.
In our analysis we decompose $w(z)$ into a sum of weighted
orthogonal step--functions to facilitate the reconstruction of
sharp features (provided binning is sufficient). Extensions of our
analysis could easily include using alterative expansion functions
to see whether other properties of the time--evolution of dark
energy are detected or constrained.

We find that the reconstruction of a dynamical $w(z)$ using
artificial datasets of $H(z)$ and $D_A(z)$ is very promising at
low redshifts but suffers from a bias towards the chosen default
model at higher $z$. To combat this effect, the prior model was
incorporated into our hypothesis space allowing the reconstruction
of a model--independent distribution of $w$, with a manageable
loss of accuracy. Once the technique was established and
demonstrated, it was applied to a combinations of the current
cosmological datasets and two popular choices of prior models,
namely a constant EOS and a mildly evolving $w(z)$.

Using a dataset including the current WMAP3 measurement of the CMB
shift parameter, the baryon acoustic oscillation measurement and
the HST Key project measurement of the Hubble parameter in
conjunction with the SN type Ia data from the SNLS project, we
found that  $w = -1$ in the redshift range $0 \leq z \leq 1100$,
with errorbars depending on the prior and model assumptions. In
the most optimistic case, where the data are supplemented by an
entropic prior around $w=-1$, the error is of order $20\%$ at all
redshifts. When the same dataset was instead supplemented by the
SN sample from the HST/GOODS program, the results agree at low
($z\lsim 0.3$) redshift. We found however that the reconstruction
tends to prefer a value $\wv
> -1$ around $z \sim 0.5$ with a significance between 1 and
2$\sigma$, depending on assumptions. This shows the dangers of
fitting a parametric form of $w(z)$ to the data, in which case one
is bound to miss possibly significant features in the
measurements. The high-redshift behaviour of the EOS becomes
increasingly dominated by the entropic prior and thus exhibits a
mild dependence on the choice of prior model. We have investigated
the correlation properties of our reconstruction, and identified a
moderate anti--correlation among the first few redshift bins of
our reconstructed points.

The MaxEnt technique presented here improves on other methods
designed to minimize noise artifacts in that the amount of
information taken from the data is not determined by the analyst
but rather dictated by the data themselves. The presence of real
structure rather than noise-induced complexity is indicated by the
size of the error bars. The entropic prior adjusts the error bars
when the information provided by the data is unreliable. Secondly,
in the absence of real information the reconstruction tends toward
our most intuitive estimate of $w(z)$ with suitably large
variance. In conclusion the merits of this technique are that it
is self--regulating in the sense that it allows the data to
determine the amount of structure that is included.  More
importantly it does not require an inherent assumption of the
functional form of the true equation of state.

We hope that this technique will prove useful in deriving even
stronger, model--independent constraints on the dark energy
history from future, high--quality data.

{\em Acknowledgments} The authors wish to thank Bruce Bassett,
Pedro Ferreira, Scott Dodelson, Josh Frieman, Dragan Huterer, Joe
Silk and Glenn Starkman for stimulating discussions and useful
suggestions. We are grateful to Sarah Bridle for suggesting to
include the model in the parameter space and to an anonymous
referee for many valuable comments. CZ is supported by a Domus A
scholarship awarded by Merton College. RT is supported by the
Royal Astronomical Society through the Sir Norman Lockyer
Fellowship and by St Anne's College, Oxford. The use of the
Glamdring cluster of Oxford University is acknowledged.


\end{document}